\documentclass{article}
\usepackage{romp}
\usepackage{amssymb}
\usepackage{xypic}
\usepackage{color}

\title{ Symmetries in Lagrangian Field Theory }
\author{ Lucia B\'{u}a \\ Departamento de Xeometr\'{\i}a e Topolox\'{\i}a
  Facultad de Matem\'{a}ticas,  USC, Spain \\ e-mail: lucia.bua@usc.es
  \\[2ex] Ioan Bucataru \\ Faculty of Mathematics,
  Alexandru Ioan Cuza University, Ia\c si, Romania
         \\ e-mail: bucataru@uaic.ro \\[3ex] Manuel de Le\'{o}n \\ ICMAT
  CSIC,  Madrid, Spain \\ e-mail: mdeleon@icmat.es \\[4ex] Modesto
  Salgado \\ Departamento de Xeometr\'{\i}a e Topolox\'{\i}a
  Facultad de Matem\'{a}ticas,  USC, Spain \\ e-mail:
  modesto.salgado@usc.es \\[4ex] Silvia Vilari\~no \\
Centro Universitario de la Defensa $\&$ I.U.M.A.,
    50090 Zaragoza, Spain \\
    e-mail: silviavf@unizar.es}

\begin{document}

\maketitle
\begin{abstract}
By generalising the cosymplectic setting for time-dependent Lagrangian mechanics,
we propose a geometric framework for the Lagrangian formulation of classical field
theories with a Lagrangian depending on the independent variables.
For that purpose we consider the first order jet bundles $J^1\pi$  of
a fiber bundle $\pi:E\to {\mathbb R}^k$ where ${\mathbb R}^k$ is the
space of independent variables. Generalized symmetries of the
Lagrangian are introduced and the corresponding Noether Theorem is
proved.
\end{abstract}


\section{Introduction}

As it is well-know, the natural arena for studying mechanics is
symplectic geometry. One interesting problem is to extend this
geometric framework for the case of classical field theories. Several
different geometric approaches are well known: the polysymplectic
formalisms developed by Sardanashvily et
al. \cite{sarda2,sarda3,sarda1}, and by Kanatchikov \cite{Kana}, as
well as the n-symplectic formalism of Norris \cite{no1,no2}, and the
$k$-cosymplectic of de Le\'on et al. \cite{mod1,mod2}.

Let us remark that the multisymplectic formalism is the most ambitious
program for developing the Classical Field Theory (see for example
\cite{CCI91, bar1,bar2,GIM1,GIM2,KijTul}, and references quoted therein).

 The aims of this paper are
\begin{itemize}
\item To give a new  Lagrangian description of first order Classical
  Field Theory, by considering  a fibration  $E\to {\mathbb R}^k$,
  which has as particular cases the cosymplectic setting for
  time-dependent Lagrangian mechanics, and     it is related with the $k$-cosymplectic
    \cite{mod2} and multisymplectic formalisms. 
    Let us observe that although every fiber bundle over $ {\mathbb R}^k$ is a trivial bundle since ${\mathbb R}^k$ is contractible (Steenrod 1951 \cite{steenrod}), we do not use this fact to develop our Lagrangian description.
  
  \item To introduce and to study the generalized symmetries in first
   order Lagrangian Field Theory. For time-dependent Lagrangian
   Mechanics this was done by J.F. Cari\~{n}ena et al. \cite{CFNM}.
\end{itemize}

In the present paper we  present a new approach for Lagrangian Field
Theory, working  with  the first  order jet bundle  $J^1\pi$  of a
fiber bundle $\pi:E\to {\mathbb R}^k$,  where   $E$ is
$(n+k)$-dimensional.

The crucial point is  that each $1$-form on ${\mathbb R}^k$ defines a tensor field of type $(1,1)$ on  $J^1\pi$, see Saunders \cite{saundersJet}.

The paper is organized as follows.  The main tools to be used are
those of vector fields, $k$-vector fields and forms along maps, the
general definitions of which are given in Section \ref{pre}.

In Section \ref{jets} we introduce the geometric elements on
$J^1\pi$ necessary to develop the geometric formulation of the
Euler-Lagrange field equations in Section \ref{lagfield} and to the
study of symmetries and  conservations laws. The principal tools, here
described, are the canonical vector fields, the $k$ vertical
endomorphisms and a kind of $k$-vector fields, known as {\sc sopde}s,
which describe systems of second order partial differential equations.

This machinery is later used to discuss symmetries in this context, extending some previous results (see \cite{BBS12,mrsv}).

The geometric formulation of the Euler-Lagrange field equations is given in Section
\ref{lagfield}, see Theorem \ref{relvectorphisol}. For this purpose we introduce the $k$ Poincar\'{e}-Cartan $1$-forms using the Lagrangian and the $k$ vertical endomorphisms.

Our formulation is a natural extension
of the   $k$-cosymplectic formalism  developed in \cite{mod2}
as we show in Section \ref{lagfield-cosy}.

Section \ref{sim} is devoted to discussing symmetries and conservation laws. We
introduce symmetries of the Lagrangian and we  give a Noether Theorem.

\section{Preliminaries}\label{pre}

\subsection{$k$-vector fields}\label{pre-k-vectors}

A system of first-order ordinary differential equations, on a
manifold $M$, can be   geometrically described as  a vector field on
$M$. Accordingly, a system of first-order partial differential
equations on $M$ can be  geometrically described as a $k$-vector
field on $M$, for some $k>1$.

Particularly,  we can identify a system of second-order partial
differential equations ({\sc sopde})  with some special $k$-vector
fields on the manifold $J^1\pi$, for some $k>1$.

We briefly recall the correspondence between systems of first-order
partial differential equations and $k$-vector fields.

Let us denote by $T^1_kM$ the Whitney sum $TM\oplus
\stackrel{k}{\dots} \oplus TM$ of $k$ copies of $TM$ and $\tau_M:T^1_kM \rightarrow M$ the canonical projection.
\begin{definition}{Definition}
 A  {\it $k$-vector field} on an arbitrary manifold $M$ is a section
 ${\bf X}: M \longrightarrow T^1_kM$ of the canonical projection
 $\tau_M:T^1_kM \rightarrow M$. \end{definition}

Each $k$-vector field ${\bf X}$ defines a family of $k$ vector fields
$X_{1}, \dots, X_{k}\in\mathfrak{X}(M)$ by projecting ${\bf X}$ onto every
factor; that is, $X_\alpha=\tau_{\alpha}\circ {\bf X}$, where
$\tau_{\alpha}\colon T^1_kM \rightarrow TM$ is the canonical
projection on the $\alpha^{th}$-copy $TM$ of $T^1_kM$.

\begin{definition}{Definition} \label{integsect}
An {\it integral section}  of the $k$-vector field ${\bf X}=(X_{1}, \dots,
X_{k})$, passing through a point $x\in M$, is a map $\psi\colon
U\subset {\mathbb R}^{k} \rightarrow M$, defined on some open
neighborhood  $U$ of $0\in {\mathbb R}^{k}$, such that
\begin{equation}\label{integcond}
\psi(0)=x, \ \psi_*(x)\left(\frac{\partial}{\partial
x^\alpha}\Big\vert_x\right)=X_{\alpha}(\psi (x))\in T_{\psi(x)}M,
\quad \forall x\in U, \ 1\leq \alpha\leq k.
  \end{equation}
 A $k$-vector field $ {\bf X}=(X_1,\ldots , X_k)$ on $M$ is said to be
{\it integrable} if there is an integral section of ${\bf X}$ passing through
every point of $M$.
\end{definition}

From Definition \ref{integsect} we deduce that $\psi$ is an integral
section of ${\bf X}=(X_{1}, \dots, X_{k})$ if, and only if, $\psi$ is a
solution to the system of first-order partial differential equations
$$
X_\alpha^i(\psi(x))= \frac{\partial \psi^i}{\partial
x^\alpha}\Big\vert_x, \quad   1\leq \alpha \leq k,  \ 1\leq i \leq
\dim M,
$$
where $X_\alpha=X^i_{\alpha} {\partial }/{\partial y^i}$ on a
coordinate system $(U,y^i)$ on  $M$, $y^i\circ \psi=\psi^i$, and
$x^{\alpha}$ are coordinates on ${\mathbb R}^k$.

For a $k$-vector field ${\bf X}=(X_1,\dots,X_k)$ on $M$ we require the integrability condition
$[X_\alpha,X_\beta]=0, \forall \alpha,\beta \in \{1,\dots,k\}$, as it
has been considered in \cite{Maranon}, see also \cite{kosambi35,kosambi48}.  
\subsection{Sections along a map} \label{pre-sections along map}

Given a fiber bundle $\pi:B\to M$ and a differentiable mapping
$f:N\to M$, {\it  a section of $\pi$ along $f$} is a differentiable
map $\sigma:N\to B$ such that $\pi\circ \sigma=f$ (see e. g.
\cite{Poor}). When $\pi$ is a vector bundle, then  the set of such
sections can be endowed with a structure of $C^\infty(N)$-module.

In the case of $\pi:B\to M$ being the tangent bundle of $M$,
$\tau_M:TM\to M$, or the cotangent bundle, $\pi_M:T^*M\to M$, the
sections along $f$ will be called {\it vector fields along $f$} and
{\it $1$-forms along $f$, respectively}.

The notion of sections along a map has been shown to be very fruitful.
Many objects commonly used in Physics find their suitable geometric
representative by means of this concept and a related one of
$f$-derivations (\cite{PidTul,saundersJet}), but they have only recently been 
introduced in Physics (\cite{CFNM91,CLM89,GraciaPons}).

Let $X$ be a vector field along $f:N\to M$, say
\[
\xymatrix@=10mm{ & TM\ar[d]^-{\tau}\\
N \ar[ru]^-{X}\ar[r]^-{f}   & M  }
\]
then we can define an $f$-derivation $i_X:\Lambda^p(M) \to \Lambda^{p-1}(N)$ of
degree $-1$  (and type $i_*$)  as follows:   $i_Xg=0$ for all $g \in C^\infty(M)$ and
\begin{equation}\label{ixalpha}
(i_X\omega)(x)\left(v_{1_x}, \ldots ,
v_{(p-1)_x}\right)=\omega\left(f(x)\right)\left(X(x),f_*(x)v_{1_x},
\ldots , f_*(x)v_{(p-1)_x}\right)
\end{equation}
where $v_{1_x}, \ldots , v_{(p-1)_x}\in T_xN$. There is another
related  $f$-derivation $d_X$ defined by
\begin{equation}\label{dd}
d_X=i_X\circ d  +d  \circ i_X,
\end{equation}
where $d$ stands for the operator of exterior differentiation. This
derivation is of degree $0$ (and type $d_*$), i.e., $d_X\circ d_{(M)}=d_{(N)}\circ
d_X$.

Note that when $X\in  \mathfrak{X}(M)$,  then the
$id_M$-derivations $i_X$ and $d_X$ are nothing but the inner product or
contraction $i_X$ and the Lie derivative $\mathcal{L}_X$,
respectively.

Let us observe that $d_{X}$ is an  $f^*$-derivation associated
to $X$ in the sense of Pidello and Tulczyjew \cite{PidTul}.

If $X$ is a vector field along $f:N\to M$, from (\ref{ixalpha}) and (\ref{dd}) we deduce that the map
$$
\begin{array}{ccccl}
d_{X}&:& C^\infty( M )& \to & C^\infty(N)\\ \noalign{\medskip} & &
F& \to &  d_{X}F
\end{array}
$$
is given by
\begin{equation}\label{dx}
(d_{X}F)(x)=(i_X\, dF)(x)=dF(f(x))(X(x))=X(x)(F),\,\,  \forall x \in N.
\end{equation}

Finally, if $\pi:B\to M$  is  a differentiable fibre bundle, we associate to
each  $\pi$-semi-basic  $p$-form $\alpha$ on $B$  a  $p$-form
$\alpha^V$ along $\pi$, as follows:
\begin{equation}\label{levvert}
\alpha^V(b)({v_1}_{\pi(b)}, \ldots,{v_p}_{\pi(b)})=
\alpha(b)({w_1}_{b}, \ldots, {w_p}_{b}), b\in B
\end{equation}
where ${v_i}_{\pi(b)}\in T_{\pi(b)}M$, $i=1, \ldots , p$, and
${w_i}_{b}\in T_{b}B$ are such that $\pi_*(b) {w_i}_{b}=
{v_i}_{\pi(b)} $.

This type of form $\alpha^V$ will be used in Section \ref{sim-gen}.

\section{The geometry of jet bundles}\label{jets}

In this paper,  we  work with  the first and second order jet
bundles $J^1\pi$ and $J^2\pi$ of a fiber bundle $\pi:E\to {\mathbb R}^k$,  where
$E$ is an $(n+k)$-dimensional manifold.

If $(x^{\alpha})$  are local  coordinates on  ${\mathbb R}^k$ and
$(x^\alpha, q^i)$ are local fiber coordinates on $E$, we consider
the induced standard jet coordinates $(x^{\alpha}, q^i, v^i_{\alpha})$ on $J^1\pi$
and  $(x^{\alpha}, q^i, v^i_{\alpha}, v^i_{\alpha\beta})$ on $J^2\pi$,
where $1\leq i\leq n$, $1\leq \alpha, \beta\leq k$. The induced jet
coordinates are given by
$$
\begin{array}{cl}
x^{\alpha}(j^1_x\phi) = x^{\alpha}(x)=x^{\alpha}, &
q^i(j^1_x\phi)=q^i(\phi(x)), \\ \noalign{\medskip}
 v^i_{\alpha}(j^1_x\phi)=\displaystyle\frac{\partial \phi^i}{\partial
   x^{\alpha}}\Big\vert_{x}, &
v^i_{\alpha\beta}(j^2_x\phi) = \displaystyle\frac{\partial^2\phi^i}{\partial
x^{\alpha}\partial x^{\beta}}\Big\vert_{x}.
\end{array}$$
Here, $\phi$ is a section for $\pi$ and $j^1_x\phi$, $j^2_x\phi$  are
the $1$-jet and $2$-jet on $x$, respectively.

For the canonical projections, we use the usual notations
$$\begin{array}{ccccccccl}
J^2\pi & \stackrel{\pi_{2,1}}{\longrightarrow} &  J^1\pi
& \stackrel{\pi_{1,0}}{\longrightarrow} & E & &  J^1\pi
& \stackrel{\pi_{1}}{\longrightarrow} & {\mathbb R}^k \\ \noalign{\medskip}
j^2_x\phi & \to & j^1_x\phi & \to & \phi(x) & & j^1_x\phi & \to & x.
\end{array}$$

\subsection{Canonical vector fields along the projections $\pi_{1,0}$
  and $\pi_{2,1}$} \label{jets-can}

The  vector field  $T^{(0)}_{\alpha}$ on $E$ along $\pi_{1,0}$ and the
vector field  $ T^{(1)}_{\alpha}$ on $J^1\pi$ along $\pi_{2,1}$
\[ \xymatrix@=10mm { & T(J^1\pi)\ar[d]^-{\tau_{J^1\pi}}& TE\ar[d]^-{\tau_E}\\
J^2\pi \ar[ru]^-{T^{(1)}_\alpha}\ar[r]^-{\pi_{2,1}}   & J^1\pi
\ar[ru]^-{T^{(0)}_\alpha}\ar[r]^-{\pi_{1,0}}   & E  }
\]
are defined respectively by
\begin{eqnarray}\label{t0t1}
T^{(0)}_{\alpha}(j^1_x\phi)=\phi_*(x) \left( \frac{\partial}{\partial
    x^{\alpha}}\Big\vert_{x}\right) \in T_{\phi(x)}E \\
\noalign{\medskip} T^{(1)}_{\alpha} (j^2_x\phi) =
(j^1\phi)_*(x) \left( \frac{\partial}{\partial
    x^{\alpha}}\Big\vert_{x} \right)\in T_{j^1_x\phi}(J^1\pi),
\end{eqnarray}
as we can see from the above diagram.

Their local expressions are given by
\begin{equation}\label{loct0t1}\begin{array}{ccl} T^{(0)}_\alpha & =
    &\displaystyle\frac{\partial}{\partial x^\alpha}\circ\pi_{1,0} + v^i_\alpha
\displaystyle\frac{\partial}{\partial q^i}\circ\pi_{1,0} \\ \noalign{\medskip} T^{(1)}_\alpha &
= &\displaystyle\frac{\partial}{\partial x^\alpha}\circ\pi_{2,1} +v^i_\alpha
\displaystyle\frac{\partial}{\partial q^i}\circ\pi_{2,1}+v^i_{\alpha\beta}
\displaystyle\frac{\partial}{\partial v^i_\beta}\circ\pi_{2,1}.
\end{array} \end{equation}

Using (\ref{dx}) we have the maps
$d_{T^{(0)}_\alpha}$ and $d_{T^{(1)}_\alpha}$
$$\begin{array}{lllll}
d_{T^{(0)}_\alpha} &:& C^\infty(E) & \longrightarrow & C^\infty( J^1\pi ) \\ \noalign{\medskip}
 d_{T^{(1)}_\alpha}&:&  C^\infty( J^1\pi ) & \longrightarrow &  C^\infty( J^2\pi )\end{array}$$
defined by $ T^{(0)}_\alpha, \, T^{(1)}_\alpha$, respectively.

From (\ref{dx}) and (\ref{loct0t1}) one obtains
\begin{equation}\label{0thetzeta40}
d_{T^{(0)}_\alpha}F=\frac{\partial F}{\partial x^\alpha}\circ\pi_{1,0} +v^i_\alpha
\, \frac{\partial F}{\partial q^i}\circ\pi_{1,0}    \, ,
\end{equation}
and
\begin{equation}\label{thetzeta4}
d_{T^{(1)}_\alpha}G=\frac{\partial G}{\partial x^\alpha}\circ\pi_{2,1} +v^i_\alpha
\, \frac{\partial G}{\partial q^i}\circ\pi_{2,1}+v^i_{\alpha\beta} \,
\frac{\partial G}{\partial v^i_\beta}\circ\pi_{2,1}
\, .
\end{equation}
where $F\in  C^\infty(E)$ and $G\in C^\infty( J^1\pi )$.

\begin{remark}{Remark}
 Let us observe that the vector fields  $   T^{(0)}_\alpha$ and $   T^{(1)}_\alpha$
 are $(\pi_{2,1},\pi_{1,0})$-related in the following sense
$(\pi_{1,0})_*\circ T^{(1)}_\alpha=T^{(0)}_\alpha\circ
\pi_{2,1}\quad .
 $
\end{remark}

\subsection{Prolongations of vector fields}\label{jets-prolong}

We recall the   prolongations of vector fields from  $E$ to $J^1\pi$  and the prolongation of a vector field along  $\pi_{1,0}$ ,
see Saunders \cite{saundersJet} (Sections $4.4$  and  Section $6.4$)  .

   Let $X$ be a vector field on $E$ locally   given by
$$ X= X_\alpha(q,v) \frac{\partial}{\partial x^\alpha}+X^i(q,v)\frac{\partial}{\partial q^i},$$  then its   prolongation $X^1$  is the vector field on $J^1\pi$
 whose  local expression is
\begin{equation}\label{locX1gen}
 X^1=X_\alpha\frac{\partial}{\partial
x^\alpha}+X^i \frac{\partial}{\partial q^i} +  \left(
 \frac{dX^i}{d x^\alpha}- v_\beta^i  \frac{dX_\beta}{d
x^\alpha}\right)  \frac{\partial}{\partial v_\alpha^i},
\end{equation}
where $d/dx^\alpha$ denotes the total derivative, that is,
$$
\frac{d}{dx^\alpha}=\frac{\partial}{\partial x^\alpha}+
v^j_\alpha\frac{\partial }{\partial q^j} \qquad  1\leq \alpha\leq k.
 $$
Let us observe that $\displaystyle\frac{dF}{dx^\alpha}=T^{(0)}_\alpha(F)$ where $F\in C^\infty(J^1\pi) $.

 Let $X$ be  a vector field along  $\pi_{1,0}$ and $X^{(1)}$  its
 first prolongation  along $\pi_{2,1}$, which means
\[\xymatrix@=10mm{ & T(J^1\pi)\ar[d]^-{\tau_{J^1\pi}}& TE\ar[d]^-{\tau_E}\\
J^2\pi \ar[ru]^-{X^{(1)}}\ar[r]^-{\pi_{2,1}}   & J^1\pi
\ar[ru]^-{X}\ar[r]^-{\pi_{1,0}}   & E  } \quad ,\]
see Saunders \cite{saundersJet}, Section $6.4$.
 If $X$ has the local expression
$$
X  =  X_\alpha(x,q,v)\, \frac{\partial}{ \partial x^\alpha}\circ \pi_{1,0}
+X^i(x,q,v) \frac{\partial }{\partial q^i}\circ\pi_{1,0}$$
then $X^{(1)}$ is locally given by
$$\begin{array}{ccl}
  X^{(1)} & =  & X_\alpha\circ\pi_{2,1} \displaystyle\frac{\partial }{\partial x^\alpha}\circ\pi_{2,1}+X^i\circ\pi_{2,1} \displaystyle\frac{\partial }{\partial q^i}\circ\pi_{2,1} \\ \noalign{\medskip}
   & & +
 \left( d_{T^{(1)}_\alpha}(X^i\circ\pi_{2,1})  - d_{T^{(1)}_\alpha}(X_\beta\circ\pi_{2,1}) v^i_\beta  \right)\frac{\partial }{\partial v^i_\alpha}\circ\pi_{2,1} \,\,\, .
\end{array}$$

If  $X$  is a vector field along  $\pi_{1,0}$ and $\pi$-vertical then it is locally given by
 \begin{equation}\label{locz1v} X^{(1)}=\left(X^i\frac{\partial }{\partial q^i}\right)\circ\pi_{2,1}+\left(\frac{ \partial X^i}{\partial x^\alpha}\circ\pi_{2,1}+ v^j_\alpha\frac{ \partial X^i}{\partial q^j}\circ\pi_{2,1} + v^j_{\alpha\beta}\frac{\partial X^i}{\partial v^j_\beta}\circ\pi_{2,1} \right)\frac{\partial }{\partial v^i_\alpha}\circ\pi_{2,1}
\end{equation}

According to  (\ref{dx}), we know that the vector field $X^{(1)}$ along $\pi_{2,1}$ defines a map
$$
\begin{array}{ccccl}
d_{X^{(1)}}&:& C^\infty( J^1\pi )& \to & C^\infty( J^2\pi )\\
\noalign{\medskip} & & F& \to &  d_{X^{(1)}}F \, \, .
\end{array}
$$
with local expression
\begin{equation}\label{thetzeta40}
d_{X^{(1)}}F=X_\alpha\circ \pi_{2,1} \frac{\partial F}{\partial x^\alpha}\circ
\pi_{2,1}+X^i\circ \pi_{2,1} \frac{\partial F}{\partial q^i}\circ \pi_{2,1}
  + \left( d_{T^{(1)}_\alpha}X^i  - v^i_\beta  d_{T^{(1)}_\alpha}X_\beta    \right)\frac{\partial F}{\partial v^i_\alpha}\circ \pi_{2,1}\, .
\end{equation}

\begin{remark}{Remark}
  Let   $X$ be a $\pi$-vertical vector field on $E$. The vector field $X\circ \pi_{1,0}$ along $\pi_{1,0}$
     \[\xymatrix@=10mm{ & TE\ar[d]\\
J^1\pi\ar[ru]^-{X\circ \pi_{1,0}}\ar[r]^-{\pi_{1,0}}   & E
}
\]
 satisfies that \begin{equation}\label{xx}
 ( X\circ \pi_{1,0})^{(1)}=X^1\circ \pi_{2,1} .
  \end{equation}
 \end{remark}

\subsection{Vertical endomorphisms}\label{jets-vert}

\bigskip

Each $1$-form $dx^\alpha, \, 1\leq \alpha\leq k$, defines a canonical tensor field $S_{ dx^\alpha}$ on $J^1\pi$ of type $(1,1)$, see  Saunders \cite{saundersJet}, page $156$,
with local expression
 \begin{equation}\label{locsalf} S_{ dx^\alpha}\equiv (dq^i-v^i_\beta dx^\beta)\otimes\frac{\partial}{\partial v^i_\alpha}\quad .
\end{equation}
Throughout the paper, we   denote $S_{ dx^\alpha}$ by $S^\alpha$.

      The vector
valued $k$-form $S$ on $J^1\pi$, defined in \cite{saunders1,saundersJet},
whose values are vertical vectors over $E$,   is given in coordinates by
\begin{equation}\label{localS}
S=\left((dq^i-v^i_\beta dx^\beta)\wedge d^{k-1}x_\alpha \right)\otimes\frac{\partial }{\partial v^i_\alpha}
\end{equation}
  where
$$ d^{k-1}x_{\alpha}=i_{{\partial}/{\partial
      x^{\alpha}}}d^kx = (-1)^{\alpha-1}dx^1\wedge \cdots \wedge dx^{\alpha-1}\wedge dx^{\alpha+1}
\wedge \cdots \wedge dx^k $$
and $d^kx=dx^1 \wedge \cdots \wedge dx^k$ is the standard volume form on $\mathbb{R}^k$.

From (\ref{locsalf}) and (\ref{localS}) we deduce that $S$ and $\{S_{dx^1},\ldots , S_{dx^k}\}$ are related by the formula
$$
S=S^\alpha \wedge d^{k-1}x_{\alpha}.
$$
We  also have
$dx^{\alpha}\wedge S=-S^\alpha \wedge d^kx.$

\subsection{Contact structures and second order partial differential equations}
\label{jets-sodes}

\bigskip
Let us consider the \emph{Cartan distribution}, which is the
$(k+nk)$-dimensional distribution, given by
$$  C(J^1\pi) = \, Ker\,  S^{1}\cup \ldots \cup \, Ker\,  S^{k}\, .
  $$

  From (\ref{locsalf}) we deduce that $X\in C(J^1\pi) $ if, and only if,
  $$ (dq^i-v^i_{\alpha}dx^{\alpha})(X)=0,$$ and thus $X$  is locally given by
$$
      X=X_{\alpha}\left(\frac{\partial}{\partial x^{\alpha}} +
        v^i_{\alpha}\frac{\partial}{\partial q^i}\right) +
      X^i_{\alpha}\frac{\partial}{\partial
        v^i_{\alpha}} .   $$
Therefore,  a local basis for $C(J^1\pi)$ is given by the local $k+nk$
vector fields
$$ \frac{\partial }{\partial x^\alpha}+v^i_\alpha\,\frac{\partial }{\partial q^i} \quad , \quad
\frac{\partial }{\partial v^i_\alpha} .$$

We also consider the
\emph{contact codistribution}, which is the $n$-dimensional distribution that
represents the annihilator of the Cartan distribution and  is given
by
\begin{eqnarray} \label{contact}
\Lambda^1_C(J^1\pi)\, =\, \{\theta \in
\Lambda^1(J^1\pi),\, \,  (j^1\phi)^*\theta=0,\, \, \forall \phi \in
\, Sec\, (\pi)\}    . \nonumber \end{eqnarray}

A local basis for $\Lambda^1_C(J^1\pi)$ is the set of canonical
$1$-forms
$$
\delta q^i=dq^i-v^i_\beta dx^\beta, \quad i=1, \ldots, n.
$$

\begin{definition}{Definition}\label{sode2}
A $k$-vector field  ${\bf \Gamma}=(\Gamma_1,\dots,\Gamma_k)$ on   $J^1\pi$
   is  said to be a second order partial differential equation  ({\sc sopde} for short)  if
$$ dx^\alpha(\Gamma_\beta)=\delta_\beta^\alpha , \quad
S^\alpha(\Gamma_\beta)=0, $$
or equivalently,
$$ dx^\alpha(\Gamma_\beta)=\delta_\beta^\alpha , \quad
 \delta q^i(\Gamma_\beta)=0\quad
$$
for all $i=1 \ldots n$, $\alpha, \beta=1 \ldots k.$
\end{definition}

 Every vector field $\Gamma_\alpha$ of  a {\sc sopde} belongs to the Cartan distribution.


 From Definition \ref{sode2} and the expression (\ref{locsalf}) of $S^\alpha$,
  we obtain  that the local
expression of a {\sc sopde} $(\Gamma_1 ,\ldots,\Gamma_k) $ is
\begin{equation}\label{localxia}
\Gamma_\alpha(x^\beta,q^i,v^i_\beta)=\frac{\partial}{\partial
x^\alpha}+v^i_\alpha\frac{\displaystyle
\partial} {\displaystyle
\partial q^i}+
\Gamma^i_{\alpha \beta} \frac{\displaystyle\partial} {\displaystyle
\partial v^i_\beta},\quad 1\leq \alpha \leq k
\end{equation}
where $\Gamma^i_{\alpha \beta} $ are functions locally defined on $J^1\pi$. As a
direct consequence of the above local expressions, we deduce that
the family of vector fields $\{\Gamma_1, \ldots , \Gamma_k\}$ are linearly
independent.

\begin{definition}{Definition}\label{de652}
 Let $\phi:{\mathbb R}^k \rightarrow E$  be  a section of $\pi$, locally given by $\phi(x^\alpha)=(x^\alpha,\phi^i(x^\alpha))$, then  the {\rm first prolongation}
$j^1\phi$ of $\phi$ is the map
\begin{equation}\label{locj1phi}
\begin{array}{rcl}
j^1\phi:{\mathbb R}^k & \longrightarrow &   J^1\pi \\
 x & \longrightarrow &
j^1_x\phi\equiv\left( x^1, \dots, x^k, \phi^i (x^1, \dots, x^k),
\frac{\displaystyle\partial\phi^i}{\displaystyle\partial x^\alpha}
(x^1, \dots, x^k)\right)
\end{array}
\end{equation} for all $\alpha=1,\ldots,k$ and for all $x\in Domain \,\phi$.
\end{definition}

We will see that the integral sections of  a {\sc sopde} are prolongations  of sections.

The following proposition has been also proved in Saunders \cite{saundersJet}.
\begin{proposition}{Proposition}\label{RelSecCont}
    A section $\psi$ of $\pi_1$ is the $1$-jet prolongation of a section of $\pi$ (in other words it is a \emph{holonomic field}) if, and only, if $\psi^*\theta=0$ for all $\theta \in \Lambda^1_c(J^1\pi)$.
\end{proposition}

Proof:
    Consider $\psi: {\mathbb R}^k \to J^1\pi$, $\psi(x^{\alpha})=(x^{\alpha}, \psi^i(x^{\alpha}), \psi^i_{\beta}(x^{\alpha}))$ a section of $\pi_1$ and $\theta=\theta_i(dq^i-v^i_{\beta}dx^{\beta}) \in \Lambda^1_c(J^1\pi)$. It follows that
    $$\psi^*\theta=\theta_i \circ \psi \left(\frac{\partial \psi^i}{\partial x^\beta} - \psi^i_\beta \right)dx^\alpha \, .$$
    Therefore, $\psi^*\theta = 0$  for all $\theta \in \Lambda^1_c(J^1\pi)$ if and only if $\psi^i_{\beta}= {\partial \psi^i}/{\partial x^{\beta}}$.
 \rule{5pt}{5pt}

Now we characterize the integral sections of  a {\sc sopde}.

\begin{proposition}{Proposition}
Let ${\bf \Gamma}$ be an integrable $k$-vector field. Then the
following  three pro\-per\-ties are equivalent:

i) ${\bf \Gamma}$ is  a {\sc sopde}

ii) The integral sections of ${\bf \Gamma}$ are $1$-jets prolongations of sections of $\pi$.

iii) There exists a section $\gamma$ for $ \pi_{2,1}$ such that $\Gamma_\alpha=T_\alpha^{(1)}\circ \gamma$.
\end{proposition}

Proof:

\framebox{i) $\Rightarrow$ ii)}
Let ${\bf \Gamma}$ be  an integrable $k$-vector field and $\psi\colon \mathbb{R}^k \to J^1\pi$ an integral section for  ${\bf \Gamma}$. Then from (\ref{integcond}) and the local expression (\ref{localxia}) of $\Gamma_\alpha$, we obtain
$$\begin{array}{ccl}
(\pi_1\circ \psi)_*(x)\left(\displaystyle\frac{\partial }{\partial x^\alpha}\Big\vert_x\right)&=
&(\pi_1)_*(\psi(x))\left(\psi_*(x)\left(\displaystyle\frac{\partial }{\partial x^\alpha}\Big\vert_x\right)\right) = (\pi_1)_*(\psi(x))\Gamma_\alpha(\psi(x))  \\ \noalign{\medskip} &=&\displaystyle\frac{\partial }{\partial x^\alpha}\Big\vert_{\pi_1(\psi(x))}
 \end{array}$$

which means
$$ \frac{\partial }{\partial x^\alpha}\Big\vert_x(x^\beta\circ\pi_1\circ\psi)= \delta^\alpha_\beta
$$
so $\pi_1\circ\psi=id_{\mathbb{R}^k }$ ,i.e., $\psi$ is a local section for $\pi_1$.

We must prove that $\psi=j^1\phi$ where $\phi$ is a section of
$\pi$. To this end we    use   Proposition \ref{RelSecCont}, showing
that $\psi^*\theta=0$ for all $\theta\in \Lambda^1_C(J^1\pi)$.

Let us assume that $\psi$ is an integral section passing through $j_x^1\varphi$, that is $\psi(x)=j_x^1\varphi$.

Now, since ${\bf \Gamma}$ is  a {\sc sopde} and $\theta\in \Lambda^1_C(J^1\pi)$, then $i_{\Gamma_\alpha}\theta=0$. Thus,

$$\begin{array}{ccl}
0&=&i_{\Gamma_\alpha}\theta(j^1_x\varphi)=\theta(j^1_x\varphi)\Gamma_\alpha(j^1_x\varphi)=\theta(j^1_x\varphi)\Gamma_\alpha(\psi(x))\\ \noalign{\medskip}
&=& \theta(\psi(x))\left( \psi_*(x)\left(\displaystyle\frac{\partial }{\partial x^\alpha}\Big\vert_x\right)\right)
=(\psi^*\theta)(x)\left(\displaystyle\frac{\partial }{\partial x^\alpha}\Big\vert_x\right)
\end{array}$$
 So, we have proved  $\psi^*\theta=0$ for all $\theta \in \Lambda^1_C(J^1\pi)$.

\framebox{ii) $\Rightarrow$ iii)}
We define
$$\begin{array}{cccl}
\gamma \colon &J^1\pi &\to &J^2\pi \\ \noalign{\medskip}
&j^1_x\sigma&\to &\gamma(j^1_x\sigma)=j^2_x\varphi,
\end{array}$$
where $j^1\varphi$ is an integral section of ${\bf \Gamma}$ passing through $j^1_x\sigma$ (i.e., $j^1_x\varphi=j^1\varphi(x)=j^1_x\sigma$).

Then
$$
(\pi_{2,1}\circ\gamma)(j^1_x\sigma)=\pi_{2,1}(\gamma(j^1_x\sigma))=\pi_{2,1}(j^2_x\varphi)=j^1_x\varphi=j^1_x\sigma
\, .$$
It follows that the map $\gamma$ so defined is a section for
$\pi_{2,1}$. Moreover, the vector field $\Gamma_\alpha$ can be expressed as $\Gamma_\alpha=T^{(1)}_\alpha\circ\gamma$, in fact,
$$
\Gamma_\alpha(j^1_x\sigma)=(j^1\varphi)_*(x)\left(\frac{\partial }{\partial x^\alpha}\Big\vert_x\right)
=T^{(1)}_\alpha(j^2_x\varphi)=(T^{(1)}_\alpha\circ\gamma)(j^1_x\sigma)\, .
$$

\framebox{iii) $\Rightarrow$ i)}
Let $\gamma$ be a section for $\pi_{2,1}$. We must prove that $\gamma$ defines  a {\sc sopde} by composition with $T^{(1)}_\alpha$. Since
$$\tau_{J^1\pi}\circ\Gamma_\alpha=\tau_{J^1\pi}\circ T^{(1)}_\alpha\circ\gamma=\pi_{2,1}\circ\gamma=id_{J^1\pi},$$
where $\tau_{J^1\pi} \colon T(J^1\pi)\to J^1\pi$ is the canonical projection, then $\Gamma_\alpha=T^{(1)}_\alpha\circ\gamma$ is a vector field on $J^1\pi$. Moreover, ${\bf \Gamma}$ is  a {\sc sopde}
$$
dx^\beta(\Gamma_\alpha)(j^1_x\sigma)=dx^\beta(T^{(1)}_\alpha\circ\gamma)(j^1_x\sigma)=\delta^\alpha_\beta
$$
and
$$\begin{array}{ccl}
\delta q^i(\Gamma_\alpha)(j^1_x\sigma)&=&\delta q^i(j^1_x\sigma)\left((T^{(1)}_\alpha\circ\gamma)(j^1_x\sigma)\right)= \delta q^i(j^1_x\sigma)\left(T^{(1)}_\alpha(j^2_x\varphi)\right)\\ \noalign{\medskip}
&=& v^i_\alpha(j^2_x\varphi)-v^i_\alpha(j^1_x\sigma)=0,
\end{array}$$
where the last identity is true because $\gamma$ is a section for $\pi_{2,1}$ so $j^1_x\sigma=(\pi_{2,1}\circ\gamma)(j^1_x\sigma)=\pi_{2,1}(j^2_x\varphi)=j^1_x\varphi$.
 \rule{5pt}{5pt}

If $j^1\phi$ is an integral section of  a {\sc sopde} ${\bf \Gamma}$, then $\phi$ is called {\it a solution of} ${\bf \Gamma}$.

From (\ref{integcond}) and (\ref{localxia}), we deduce
 that $\phi$ is a solution of ${\bf \Gamma}$ if, and only if,
  $q^i\circ \phi=\phi^i$ is a solution to
  the following system of second order partial differential equations
\begin{equation}\label{xisol}
 \frac{\partial^2 \phi^i}
{\partial x^\alpha \partial x^\beta}\Big\vert_x=\Gamma^i_{\alpha
  \beta} \left(x,\phi^i(x),\frac{\partial \phi^i}{\partial x^\gamma}\right).
\end{equation}
where $1\leq i\leq n$ and $1\leq \alpha,\beta \leq k$.

 The integrability conditions for the system (\ref{xisol}) requires
that the $k$-dimensional distribution induced by the {\sc sopde}
${\bf \Gamma} $, $H_0=\, span\, \{\Gamma_1,..., \Gamma_k\}$ is
integrable. The local expression (\ref{localxia}) for  a {\sc sopde}
${\bf \Gamma}$ shows that $[\Gamma_{\alpha},
\Gamma_{\beta}]=A^{\gamma}_{\alpha\beta}\Gamma_{\gamma}$ if, and only if,
$A^{\gamma}_{\alpha\beta}=0$. Therefore, for  a {\sc sopde} ${\bf \Gamma}$, we assume throughout the
paper the following integrability conditions
\begin{eqnarray}
\Gamma^i_{\alpha\beta}=\Gamma^i_{\beta\alpha}, \quad
\Gamma_{\alpha}(\Gamma^i_{\beta\gamma})=\Gamma_{\beta}(\Gamma^i_{\alpha\gamma}), \label{symxi}
\end{eqnarray}
which are equivalent to the condition $[\Gamma_{\alpha}, \Gamma_{\beta}]=0$
for all $\alpha,\beta=1, \ldots, k$.

\bigskip

\section{LAGRANGIAN FIELD THEORY}\label{lagfield}

\subsection{Poincar\`{e}-Cartan 1-forms}\label{lagfield-pc forms}

\bigskip

Let  $L:J^1\pi \to \mathbb{R}$ be a Lagrangian.

For each $\alpha=1 \ldots k$ we define the  {\it Poincar\`{e}-Cartan 1-forms} $\Theta_L^\alpha$ on $J^1\pi$ as the $1$-forms
$$\Theta^{\alpha}_{L}= Ldx^{\alpha} +
  dL\circ S^{\alpha} \, , \quad  1\leq \alpha\leq k .
 $$
Their local expressions are given by
  \begin{equation}\label{thetal}
  \Theta^\alpha_L= \left(L\delta^{\alpha}_{\beta} -\frac{\partial L}{\partial
      v^i_{\alpha}}v^i_{\beta}\right) dx^{\beta} +
  \frac{\partial L}{\partial v^i_{\alpha}} dq^i,
  \end{equation}
  or equivalently
  \begin{equation}\label{thetal1}
  \Theta^\alpha_L= \
  \frac{\partial L}{\partial v^i_{\alpha}} (dq^i-v^i_\beta dx^\beta)+Ldx^\alpha= \frac{\partial L}{\partial v^i_{\alpha}}\delta q^i+Ldx^\alpha\,\, .
  \end{equation}
Let us observe that the fact of working with a fiber bundle over  ${\mathbb R}^k$ allows us
to introduce the tensors $S^\alpha$ and consequently the  $1$-forms $\Theta^\alpha_L$.

It is known that a Lagrangian $L$ induces a $k$-form $\Theta_L$,
called the Cartan form in first-order field theories, Saunders \cite{saunders1,saundersJet},
  which is given by
$$
\Theta_L=Ld^kx+ dL\circ S. \label{tl} $$
with local expression
$$
  \Theta_L= \frac{\partial L}{\partial v^i_{\alpha}} (dq^i-v^i_\beta dx^\beta)\wedge(i_{\frac{\partial }{\partial x^\alpha}}\wedge d^kx)+Ld^kx \, \, .
  $$

 The relationship between the Cartan form and the Poincar\`{e}-Cartan 1-forms
 is given by the following identity
\begin{equation}\label{relat}
\Theta_L =\Theta_L ^\alpha \wedge d^{k-1}x_\alpha+ (1-k)L d^kx\quad .
\end{equation}

 \subsection{Euler-Lagrange field equations }\label{lagfield-EL}

 \bigskip

Let ${\bf \Gamma}=(\Gamma_1,\dots,\Gamma_k)$ be an integrable {\sc sopde}. A  direct computation, using (\ref{localxia}), (\ref{symxi}) and (\ref{thetal1}), gives the formula

\begin{equation}\label{lthetl}
\mathcal{L}_{\Gamma_\alpha}\Theta^{\alpha}_L=dL+
\left(\Gamma_\alpha \left( \frac{\partial L}{\partial v^i_\alpha} \right) - \frac{\partial L}{\partial q^i}\right)
(dq^i-v^i_\beta dx^\beta) ,
\end{equation}
and proves the following lemma.
\begin{lemma}{Lemma}
Let ${\bf \Gamma}=(\Gamma_1,\dots,\Gamma_k)$ be an integrable {\sc sopde} satisfying

 $$
 \Gamma_\alpha \left(\frac{\partial L}{\partial v^i_\alpha}\right)-\frac{\partial L}{\partial q^i} =0\,,\quad i=1,\ldots , n\, .
$$
If  $j^1\phi$  is an integral section of ${\bf \Gamma}$, then $\phi$ is
a solution to the Euler -Lagrange equations
\begin{equation}\label{e-l-2}
\frac{\partial^2 L}{\partial x^\alpha \partial v^i_\alpha}\Big\vert_{j^1_x\phi}+   \frac{\partial \phi^j}{\partial x^\alpha}\Big\vert_{x}
\frac{\partial^2 L}{\partial q^j \partial v^i_\alpha}\Big\vert_{j^1_x\phi}+   \frac{\partial^2 \phi^j}{\partial x^\alpha \partial x^\beta}\Big\vert_{x}  \frac{\partial^2 L}{\partial v^j_\beta \partial v^i_\alpha}\Big\vert_{j^1_x\phi}=
\frac{\partial  L}{\partial q^i}\Big\vert_{j^1_x\phi},
\end{equation}
 where $x\in Domain \, \phi$. Equations (\ref{e-l-2}) are  usually written as follows
$$ \frac{\partial }{\partial x^\alpha}\Big\vert_x\left(\frac{\partial L}{\partial v^i_\alpha}\circ
  j^1\phi \right)=\frac{\partial L}{\partial q^i}\Big\vert_{j^1_x\phi}, \quad 1\leq i\leq k \, .$$
\end{lemma}
\bigskip

From (\ref{lthetl}) we deduce the following proposition:

\begin{proposition}{Proposition}\label{solELvect}
Let ${\bf \Gamma}$ be an integrable {\sc sopde}, then
\begin{equation}\label{lthetl1}
\mathcal{L}_{\Gamma_\alpha}\Theta^{\alpha}_L=dL
\end{equation}
if and only if
$$
 \Gamma_\alpha \left(\frac{\partial L}{\partial v^i_\alpha}\right) - \frac{\partial L}{\partial q^i}
 =0\,,\quad i=1, \ldots, n.
$$
\end{proposition}

Thus, we have the following result
\begin{corollary}{Corollary} Let ${\bf \Gamma}$ be an integrable {\sc sopde};
 if $j^1\phi$ is solution to (\ref{lthetl1}), that is $ \mathcal{L}_{\Gamma_\alpha}\Theta^{\alpha}_L(j^1\phi)=dL(j^1\phi)$,  then $\phi$ is solution to the Euler-Lagrange equations.
\end{corollary}

\bigskip

From (\ref{lthetl1}) and the identity $i_{\Gamma_\alpha}\Theta^\alpha_L=kL$ we deduce the following proposition:

 \begin{proposition}{Proposition}\label{relsopdeELeq} Let ${\bf \Gamma}$ be  a {\sc sopde}, then
   the equation (\ref{lthetl1}) is equivalent to the equation
     $$
i_{\Gamma_\alpha} \Omega^\alpha_L=(k-1)dL \, ,
$$
where $\Omega^\alpha_L=-d\Theta^\alpha_L$   will be  called the Poincar\'{e}-Cartan $2$-forms.
\end{proposition}

\bigskip
Taking into account the above results we are able  to prove a theorem that  gives us a new Lagrangian field formulation for a bundle $E\to \mathbb{R}^k $.

First we recall that a Lagrangian $L:J^1\pi \to \mathbb{R}$ is said to be regular if the matrix
$$\left(\frac{\partial^2L}{\partial v^j_\alpha \partial v^i_\beta}\right)\quad 1\leq\, \alpha,\beta\, \leq k, \, \,\, 1\leq\, j,i\, \leq n$$
is not singular, where $1\leq\alpha,\beta\leq k, \, 1\leq j,i\leq n$.
\begin{theorem}{Theorem}\label{relvectorphisol}
Let ${\bf X}=(X_1, \ldots, X_k)$ be a $k$-vector on $J^1\pi$ such that
\begin{equation}\label{geoverel}
dx^\alpha(X_\beta)= \delta^\alpha_\beta \quad , \quad i_{X_\alpha} \Omega^\alpha_L=(k-1)dL \, ,
\end{equation}

\begin{enumerate}
\item
If $L$ is regular then ${\bf X}=(X_1, \ldots, X_k)$ is  a {\sc sopde}. If ${\bf X}$ is integrable and $\phi\colon U_0 \subset \mathbb{R}^k  \to E$ is a solution of ${\bf X}$ then $\phi$ is a solution to the Euler-Lagrange equations (\ref{e-l-2}).

\item
Now, if ${\bf X}=(X_1, \ldots, X_k)$ is integrable and  $j^1\phi$ is an integral section  of ${\bf X}$,  then  $\phi$ is a solution to the Euler-Lagrange equations (\ref{e-l-2}).

\end{enumerate}

\end{theorem}

Proof:   $(i)$ If   we write each $X_\alpha$ in local coordinates as
 $$ X_\alpha  = \frac{\partial}{\partial x^\alpha}+X^i_\alpha\frac{\displaystyle
\partial} {\displaystyle \partial q^i}+
X^i_{\alpha \beta} \frac{\displaystyle\partial} {\displaystyle \partial v^i_\beta},\quad 1\leq \alpha \leq k, $$
then, from (\ref{thetal}) and (\ref{geoverel}),  we obtain
$$\begin{array}{cccl}
(1) &  X_\alpha \left (L \delta ^\alpha_\beta - \displaystyle\frac{\partial L}{\partial
  v^i_{\alpha}} v^i_{\beta}\right) + (v^i_{\alpha} -X^i_{\alpha})
\displaystyle\frac{\partial^2 L}{\partial x^\beta \partial v^i_\alpha} & = & \displaystyle\frac{\partial L}{\partial x^\beta} \\ \noalign{\medskip}
(2) & X_\alpha \left(\displaystyle\frac{\partial L}{\partial v^j_\alpha}\right) +  (v^i_\alpha - X^i_{\alpha})
\displaystyle\frac{\partial^2 L}{\partial q^j \partial v^i_\alpha} &=&
\displaystyle\frac{\partial L}{\partial q^j} \\ \noalign{\medskip}
(3) & (v^i_\alpha - X^i_\alpha) \displaystyle\frac{\partial^2L}{\partial v^j_\alpha \partial v^i_\beta}&=& 0 .
\end{array}$$
Using that $L$ is regular we deduce from $(3)$ that $X^i_\alpha= v^i_\alpha$.

Since $\mathbf{X}$ is  a {\sc sopde} and we assume it to be integrable, its integral sections are prolongations $j^1\phi$ of sections $\phi$ of $\pi$. Then, from
$$X_\alpha(j^1\phi_x)=(j^1\phi)_*(x)
\left(\frac{\partial }{\partial x^\alpha}\Big\vert_{x}\right) $$
we deduce \begin{equation}\label{tec1}
X^i_\alpha\circ j^1\phi=v^i_\alpha\circ j^1\phi= \frac{\partial \phi^i}{\partial x^\alpha}\, , \quad X^i_{\alpha\beta}\circ j^1\phi =\frac{\partial^2\phi^i}{\partial x^\alpha x^\beta}\, ,
\end{equation}
and from $(2)$,  we obtain

$$
\frac{\partial^2 L}{\partial x^\alpha \partial v^i_\alpha} +    \frac{\partial \phi^j}{\partial x^\alpha}
\frac{\partial^2 L}{\partial q^j \partial v^i_\alpha} +  \frac{\partial^2\phi^j}{\partial x^\alpha \partial x^\beta} \frac{\partial^2 L}{\partial v^j_\beta \partial v^i_\alpha} =
\frac{\partial  L}{\partial q^i}.
$$
which means that  $ \phi$   is solution to the Euler-Lagrange
equations  (\ref{e-l-2}).

$(ii)$ If $j^1\phi$ is an integral section of ${\bf X}$, then from the local expression (\ref{locj1phi}) of $j^1\phi$ we obtain the equations (\ref{tec1}). Thus, from equation $(2)$ and  (\ref{tec1}) we deduce that $\phi$ is a solution to the Euler-Lagrange equations (\ref{e-l-2}).
 \rule{5pt}{5pt}

\begin{remark}{Remark}
  For   $k=1$, that is the fibration is $E\to \mathbb{R}$,  we obtain the well-known dynamical equation $i_\Gamma d\Theta=0$,  where ${\bf \Gamma}$ is a   second order differential equation {\sc sode}, see \cite{CFN1993}.

 In the case $E=\mathbb{R}\times Q\to \mathbb{R}$, equations (\ref{geoverel}) can be written as
the dynamical equations
$$dt(X)=1, \quad i_X \Omega_L=0 $$
where $\Omega_L$ is the Poincar\'e-Cartan $2$-form defined by the Lagrangian $L\colon \mathbb{R}\times TQ\to \mathbb{R}$. These equations coincide with  the cosymplectic formulation of the non-autonomous mechanics, see \cite{clm,lr}.

 The relationship between this formalism and the $k$-cosymplectic formalism, described in \cite{mod2} (see also \cite{EMR,krupkova97, msv,RRSV11}), will be analyzed in the  next section.  
\end{remark}

\begin{example}{Example} The Klein-Gordon equation
The equation of a scalar field $\phi$ (for instance the gravitational field) which acts on the
four-dimensional space-time is \cite{KijTul}
\begin{equation}\label{scalar}
(\Box+m^2)\phi=F'(\phi)
\end{equation}
where $m$ is the mass of the particle over which the field acts, $F$ is a scalar function such that
$F(\phi)-\frac{1}{2}m^2\phi^2$ is the potential energy of the particle of mass $m$, and $\Box$ is
the Laplace-Beltrami operator given by
\[
\Box\phi:={\rm div}\,{\rm grad} \phi=\frac{1}{\sqrt{-g}}\frac{\partial}{\partial x^\alpha}(\sqrt{-g}g^
{\alpha\beta}\frac{\partial\phi}{\partial x^\beta})\,,
\]  $(g_{\alpha\beta})$ being a pseudo-riemannian metric tensor in the four-dimensional space-time of signature
$(-+++)$.

Now we consider the trivial bundle $\pi:E=\mathbb{R}^4\times \mathbb{R} \to \mathbb{R}^4$,
with coordinates $(x^1,\ldots x^4,q)$ on $E$, and $(x^1,\ldots
x^4,q,v_1,\ldots,v_4)$ the induced coordinates on $ J^1\pi$.

 Let $L$ be the Lagrangian
$L:J^1\pi \to \mathbb{R}$ defined by
$$
L(x^1,\ldots x^4,q, v_1,\ldots, v_4)=\sqrt{-g}(F(q)-\frac{1}{2}m^2q^2+\frac{1}{2}g^{\alpha\beta}v_\alpha v_\beta),
$$
 which is regular.

Let us assume that ${\bf X}=(X_1,X_2,X_3,X_4)$ is an integrable  $4$-vector field on $J^1\pi $ solution to the equations (\ref{geoverel}), that is
\begin{equation}\label{sc}
dx^\alpha(X_\beta)=\delta^\alpha_\beta \,\, ,\quad i_{X_1}\Omega^1_L+ i_{X_2}\Omega^2_L +i_{X_3}\Omega^3_L+ i_{X_4}\Omega^4_L = 3dL
\end{equation}
From (\ref{sc}) we deduce
$$
X_\alpha\left(\frac{\partial L}{\partial v_\alpha}\right)=\frac{\partial L}{\partial q}, $$ which is
equivalent to
$$X_\alpha
( \sqrt{-g}g^{\alpha\beta}v_\beta )=\frac{\partial L}{\partial q}=\sqrt{-g}(F'(q)-m^2q)\, .
 $$
Since $L$ is regular, then  ${\bf X}$ is  a {\sc sopde}, and if $j^1\psi$ is an integral section of ${\bf X}$, then
$$(j^1\psi)_*(x)\left(\frac{\partial }{\partial x^\alpha}\Big\vert_x\right)\left(\sqrt{-g}g^{\alpha\beta}v_\beta \right)= \sqrt{-g}(F'(\psi(x))-m^2\psi(x))
$$
which can be written as
\[ 0=\sqrt{-g}\frac{\partial }{\partial x^\alpha}\left(g^{\alpha\beta}
\frac{\partial \psi}{\partial x^\beta}\right)-\sqrt{-g}(F'(\psi)-m^2\psi) \] and thus we obtain that $\psi$
is a solution to the scalar field equation (\ref{scalar}).
\end{example}

\begin{remark}{Remark}  Some particular cases of the scalar field equation (\ref{scalar}) are:
\begin{enumerate}
\item If $F=0$, we obtain the linear scalar field equation.
\item If $F(q)=m^2q^2$,  we obtain the Klein-Gordon equation, see \cite{saletan},
$$(\Box+m^2)\phi=0\,.$$
\end{enumerate}
\end{remark}

\subsection{ Relationship with the $k$-cosymplectic formalism}\label{lagfield-cosy}

      We shall show the difference between the corresponding Poincar\'e-Cartan  forms in this
new formalism and the Poincar\'e-Cartan  forms of the k-cosymplectic formalism. Let us
remark that this new approach is closer to the multysimplectic description, see identity
(\ref{relat}), than to the k-cosymplectic one.

Throught this  section we consider  the trivial bundle $E=\mathbb{R}^k \times Q \to \mathbb{R}^k $.
In this case, the manifold $J^1\pi$ of 1-jets of sections of the trivial
bundle $\pi:\mathbb{R}^k  \times Q \to \mathbb{R}^k $ is diffeomorphic to $\mathbb{R}^k
\times  T^1_kQ$, where $T^1_kQ$ denotes the Whitney sum $TQ\oplus\stackrel{k}{\ldots}\oplus TQ$ of $k$ copies of $TQ$.

The diffeomorphism is  given by
$$
\begin{array}{rcl}
J^1\pi  & \to & \mathbb{R}^k    \times T^1_kQ \\
\noalign{\medskip} j^1_x\phi= j^1_x(Id_{\mathbb{R}^k },\phi_Q) & \to & (
x,v_1, \ldots ,v_k)
\end{array}
$$
where $\phi_Q: \mathbb{R}^k  \stackrel{\phi}{\to}  \mathbb{R}^k \times Q \stackrel{\pi_Q}{\to}Q
$, and
$$
v_\alpha=(\phi_Q)_*(x)\left(\frac{\partial}{\partial
x^\alpha}\Big\vert_x\right)\, , \quad  1\leq \alpha \leq k \, .
$$

Now we recall the $k$-cosymplectic Lagrangian formalism, beginning with the necessary geometric elements

\begin{itemize}
\item

The Liouville vector field $\Delta$ on $\mathbb{R}^k \times T^1_kQ$  is the infinitesimal generator of the following flow
$$\begin{array}{ccl}
\mathbb{R} \times(\mathbb{R}^k\times T^1_kQ) & \to & \mathbb{R}^k\times T^1_kQ \\ \noalign{\medskip}
(s,(t,v_{1_q},\cdots,v_{k_q})) & \to & (t, e^sv_{1_q},\cdots,e^sv_{k_q})
\end{array}$$
and in local coordinates it has the form
$\Delta=v^i_\alpha \displaystyle\frac{\partial }{\partial v^i_\alpha} \, .$

\item The canonical vector field  $\Delta^\alpha_\beta\,,\,\, 1\leq \alpha,\beta \leq k$ is the vector field on
$\mathbb{R}^k \times T^1_kQ$  defined by
$$
\Delta^\alpha_\beta(x,{v_1}_q,\ldots , {v_k}_q)=
\frac{d}{ds}\Big\vert_{0}\left(x,{v_1}_q, \ldots, {v_{\alpha-1}}_q, {v_\alpha}_q+s{v_\beta}_q,{v_{\alpha+1}}_q,\ldots, {v_{k}}_q\right)
$$
with local expression $\Delta^\alpha_\beta=v^i_\beta \displaystyle\frac{\partial }{\partial v^i_\alpha}$. Let us observe that $\Delta=\Delta^\alpha_\alpha$.

\item For a Lagrangian $L:\mathbb{R}^k \times T^1_kQ \to \mathbb{R}$, the {\it energy  function} is defined as $E_L=\Delta(L)-L $  and its  local expression is
\begin{equation}\label{LagranEner}
E_L=v^i_\alpha\frac{\partial L}{\partial v^i_\alpha}-L \quad .
\end{equation}

\item  The canonical $k$-tangent structure on $T^1_kQ$ is the family $\{J^1,\ldots, J^k\} $ of tensor fields locally given by
$$J_\alpha= \frac{\partial }{\partial v^i_\alpha}\otimes dq^i \, .$$

   The {\it natural extension} $J^\alpha$ of the tensor fields  $J^\alpha$ on $T^1_kQ$ to $\mathbb{R}^k\times T^1_kQ$ will be denoted by $\widetilde{J}^\alpha$.

   \item The
     Poincar\'e-Cartan $1$-forms introduced in \cite{mod2}    are defined as  follows
$$\theta_L^\alpha=dL \circ  \widetilde{J}^\alpha  \quad    1\leq
\alpha \leq k \, , $$
and they have the local expression
\begin{equation}\label{am20}
\theta_L^\alpha=   \frac{\displaystyle\partial
L}{\displaystyle\partial v^i_\alpha} \,  dq^i\,.
\end{equation}
The corresponding Poincar\'e-Cartan $2$-forms  are $\omega^\alpha_L=-d\theta_L^\alpha$. From  (\ref{thetal}) and (\ref{am20}) we deduce that the relationship between the Poincar\'e-Cartan $1$-forms $\Theta^\alpha_L$ and $\theta^\alpha_L$ is given by the following equation
\begin{equation}\label{am21}\Theta^\alpha_L=\theta^\alpha_L+\left(\delta^\alpha_\beta L -\Delta^\alpha_\beta(L)\right)dx^\beta .\end{equation}

\end{itemize}

As a consequence of (\ref{am21}),  the solutions $(X_1, \ldots,X_k)$ of  our geometric field equations
$$dx^\alpha(X_\beta)= \delta^\alpha_\beta \quad , \quad i_{X_\alpha} \Omega^\alpha_L=(k-1)dL \,   $$
coincide with the solutions of   the $k$-cosymplectic field equations
$$
\begin{array}{l}
dx^\alpha(X_\beta)= \delta^\alpha_\beta, \quad 1 \leq \alpha , \beta \leq k\, ,\\
\noalign{\medskip} i_{X_\alpha} \omega_L^\alpha   =
dE_L + \displaystyle\frac{\partial L}{\partial x^\alpha}dx^\alpha \, .
\end{array}
$$
introduced in \cite{mod2}, and also the corresponding integral sections, if they exist.

\section{SYMMETRIES AND CONSERVATION LAWS}\label{sim}

The set of $k$-vector fields  solution to the equation
(\ref{geoverel})  will be denoted by $\mathfrak{X}^k_L(J^1\pi)$. As a  consequence of Propositions \ref{solELvect}  and \ref{relsopdeELeq} we have that
    an integrable {\sc sopde} ${\bf \Gamma}$ belongs to   $\mathfrak{X}^k_L(J^1\pi)$ if, and only if, $\mathcal{L}_{\Gamma_\alpha}\Theta^{\alpha}_L=dL$.

\begin{definition}{Definition}
 A {\rm conservation law} (or a {\rm conserved quantity})
for the Euler-Lagrange equations (\ref{e-l-2})
is a map  ${\it G}=(G^1 , \ldots , G^k)\colon
J^1\pi \to  {\mathbb R}^k$ such that the divergence of
$$
{\it G}\circ j^1\phi=(G^1 \circ j^1\phi, \ldots , G^k \circ
j^1\phi)\colon U\subset{\mathbb R}^k \to  {\mathbb R}^k
$$
is zero for every section  $\phi:U\subset \mathbb{R}^k  \to E$,
solution to  to the Euler-Lagrange equations (\ref{e-l-2}), which
means that for all $x\in U\subset {\mathbb R}^k$ we have
$$
0 = [Div({\it G}\circ j^1\phi)](x)=
   \frac{\partial (G^\alpha \circ j^1\phi)}{\partial x^\alpha}\Big\vert_{x}=
  j^1\phi_*(x)\Big(\frac{\partial}{\partial x^\alpha}\Big\vert_{x}\Big)(G^\alpha)=T^{(1)}_\alpha(j^2_x\phi)(G^\alpha).
$$
\end{definition}

We can characterize conservation laws for the Euler-Lagrange equations in terms of the SOPDEs in $\mathfrak{X}^k_L(J^1\pi)$.

\begin{proposition}{Proposition}\label{caractconslaw}
The map ${\it G}=(G^1 , \ldots , G^k)\colon J^1\pi \to  {\mathbb R}^k$ defines
a conservation law for the Euler-Lagrange equations (\ref{e-l-2}) if,
and only if, for every integrable  {\sc sopde} $ {\bf
  \Gamma}=(\Gamma_1,\dots,\Gamma_k)\in \mathfrak{X}^k_L(J^1\pi)$  we
have that
$${\mathcal L}_{\Gamma_\alpha}G^\alpha=0.$$
\end{proposition}

Proof:
Let $j^1_x\phi$ be an arbitrary point of $J^1\pi$. Since ${\bf \Gamma}$ is an integrable  {\sc sopde}, let us denote by $j^1\psi$ the integral section of ${\bf \Gamma}$ passing through by $j^1_x\phi$, which means
$$j^1\psi(0)=j^1_x\psi=j^1_x\phi, \quad
\Gamma_\alpha(j^1_x\psi)=(j^1\psi)_*(x)\left(\frac{\partial }{\partial x^\alpha}\Big\vert_x\right),
\quad x\in \mbox{Domain}\, \psi\,.$$

Since ${\bf \Gamma} \in \mathfrak{X}^k_L(J^1\pi)$, and $\psi$ is an integral section of ${\bf \Gamma}$ then $\psi$ is a solution to the Euler-Lagrange equations (\ref{e-l-2}). As $ {\it G}=(G^1 , \ldots , G^k)$ is a conservation law, then by hypothesis
$$\frac{\partial (G^\alpha \circ j^1\psi)}{\partial x^\alpha}\Big\vert_{0}=0, $$
and therefore we deduce
$$
{\mathcal L}_{\Gamma_\alpha}G^\alpha(j^1_x\phi)=\Gamma_\alpha(j^1_0\psi)(G^\alpha)=(j^1\psi)_*(0) \left(\frac{\partial }{\partial x^\alpha}\Big\vert_0\right)(G^\alpha)= \frac{\partial (G^\alpha \circ j^1\psi)}{\partial x^\alpha}\Big\vert_{0}=0.
$$

Conversely, we must prove that
$$\frac{\partial (G^\alpha \circ j^1\phi)}{\partial x^\alpha}\Big\vert_{x}=0, $$
for all sections  $\phi:W\subset\mathbb{R}^k \to E$, which are solutions to the  Euler-Lagrange equations (\ref{e-l-2}).

Since $j^1\phi\Big\vert_{W} :W\subset\mathbb{R}^k  \to J^1\pi$ is an injective immersion ($j^1\phi$ is a section and hence its image is an embedded submanifold), we can define a k-vector field  ${\bf X}= (X_1,\ldots, X_k)$ in $j^1\phi(W)$ as follows:
$$X_\alpha(j^1_x\phi)=(j^1\phi)_*(x)\left(\frac{\partial }{\partial x^\alpha}\Big\vert_x\right)=\frac{\partial }{\partial x^\alpha}\Big\vert_{j^1_x\phi}+\frac{\partial\phi^i}{\partial x^\alpha}\frac{\partial }{\partial q^i}\Big\vert_{j^1_x\phi}+\frac{\partial^2\phi^i}{\partial x^\alpha\partial x^\beta}\frac{\partial }{\partial v^i_\beta}\Big\vert_{j^1_x\phi}$$
  and so $j^1\phi $ is an integral section of $\mathbf{X}$  and $\mathbf{X}$ it is an integrable  {\sc sopde} on $j^1\phi(W)$.

Now we  prove that $\mathbf{X}\in \mathfrak{X}^k_L(j^1\phi(W))$. A direct computation shows that
$$
\left(X_\alpha\left(\frac{\partial L}{\partial v^i_\alpha}\right)
-\frac{\partial L}{\partial q^i} \right)\Big\vert_{j^1\phi(W)}=0
 $$
Now, since $\mathbf{X}$ is an integrable  {\sc sopde}, from Proposition \ref{solELvect} and Proposition \ref{relsopdeELeq} we deduce that  $\mathbf{X}$ is a solution to the equations (\ref{geoverel}),  and then $\mathbf{X}\in \mathfrak{X}^k_L(j^1\phi(W))$.

The following identities finish the proof:
$$
\frac{\partial (G^\alpha \circ j^1\phi)}{\partial x^\alpha}\Big\vert_{x}=(j^1\phi)_*(x)\left(\frac{\partial }{\partial x^\alpha}
\Big\vert_x\right)(G^\alpha)=X_\alpha(j^1_x\phi)(G^\alpha)=
{\mathcal L}_{X_\alpha}G^\alpha(j^1_x\phi)=0.
$$
 \rule{5pt}{5pt}

 \subsection{Generalized symmetries. Noether's Theorem}\label{sim-gen}

 \bigskip
 In this section, we introduce the (generalized) symmetries of the Lagrangian and we prove a Noether's theorem which associates to each symmetry a conservation law.

The following proposition can be seen as  a motivation of the condition  (\ref{luc00}) in the definition of generalized symmetry, and it is also a generalization of  Proposition 3.15 in \cite{rsv07}.
\begin{proposition}{Proposition}\label{600}   Let  $X$ be a   $\pi$-vertical vector field on $E$.
If there exist functions $g^\alpha:E\to \mathbb{R} \,\,, 1\leq \alpha\leq k$  such that $$X^1(L) = d_{T^{(0)}_\alpha}g^\alpha $$    then the
functions $\,\, G^\alpha=(\pi_{1,0})^*g^\alpha-\Theta^\alpha_L(X^1)   \,\,$
define a conservation law.
\end{proposition}

Proof:
Let us observe that locally
$$
 G^\alpha=g^\alpha\circ \pi_{1,0}-(X^i\circ \pi_{1,0})\frac{\partial L}{\partial v^i_\alpha}\, .
$$
Then, taking into account (\ref{0thetzeta40}) and (\ref{locX1gen}), we deduce that for every solution $\phi$ of the Euler-Lagrange equations (\ref{e-l-2}) we have
$$
\frac{\partial (G^\alpha \circ j^1\phi)}{\partial x^\alpha}\Big\vert_x=
\frac{\partial }{\partial x^\alpha}\Big\vert_x\left(g^\alpha\circ \phi - (X^i\circ \phi) \,\,(\frac{\partial L}{\partial v^i_\alpha}\circ j^1\phi)   \right)
=   [  d_{T^{(0)}_\alpha} g^\alpha - X^1(L)](j^1_x\phi)= 0
   $$
so, $(G^1,\dots, G^k)$ defines a conservation law. 
 \rule{5pt}{5pt}

The Euler-Lagrange form $\delta L$ is the $1$-form on $J^2\pi$ given by
$$
\delta L=d_{T^{(1)}_\alpha}\Theta_L^\alpha - \pi_{2,1}^*dL
$$
with local expression
$$
\delta L=\left( T^{(1)}_\alpha \left(\frac{\partial L}{\partial v^i_\alpha}\right) -
  \frac{\partial L}{\partial q^i}\circ \pi_{2,1}\right) (dq^i-v^i_\beta dx^\beta).
$$
This is a $\pi_{2,0}$-semi-basic form,
and we consider its associated form $(\delta L)^V$ along $\pi_{2,0}$, see
(\ref{levvert}), with local expression
\begin{equation}\label{locdeltal}
\left( T^{(1)}_\alpha(\frac{\partial L}{\partial v^i_\alpha})-\frac{\partial L}{\partial q^i}\circ \pi_{2,1}\right)
(dq^i\circ \pi_{2,0} -v^i_\beta dx^\beta\circ \pi_{2,0})  \, .
\end{equation}

The $1$-forms   $\Theta^{\alpha}_{L}$ are $\pi_{1,0}$-semi-basic $1$-form and   its associated forms $(\Theta_L^\alpha)^V$ along $\pi_{1,0}$ are locally given by
\begin{equation}\label{locthetahat}
  (\Theta_L^\alpha)^V=(L\delta^{\alpha}_{\beta} -\frac{\partial
L}{\partial
      v^i_{\alpha}}v^i_{\beta})\, dx^\beta \circ \pi_{1,0} +\frac{\partial L}{\partial v^i_\alpha} \, dq^i\circ \pi_{1,0} \, .
\end{equation}

 The following lemma will be useful in the study of generalized symmetries.
 \begin{lemma}{Lemma}\label{lemsimgen} Let $X$ be a
   $\pi$-vertical vector field  along $\pi_{1,0}$. Then
\begin{enumerate}
    \item If there exists functions $G^\alpha:J^1\pi \to \mathbb{R}$, $\alpha =1 ,\ldots , k$, such that
$$
   d_{T_\alpha^{(1)}}G^\alpha(j^1\phi)=-(\delta L)^V(X\circ \pi_{2,1})(j^1\phi)
$$ for any $ \phi$ solution to the Euler-Lagrange equations,
then $(G^1, \ldots , G^k)$ is a conservation law.

\medskip

\item  The following identity holds
\begin{equation}\label{iden00}
d_{X^{(1)}}L=-(\delta L)^V(X\circ \pi_{2,1})+ d_{T_\alpha^{(1)}}
\left[ (\Theta_L^\alpha)^V(X)   \right] \, .
\end{equation}

\end{enumerate}
 \end{lemma}

Proof:  \
\begin{enumerate}

\item

From (\ref{t0t1})   we have
\begin{equation}\label{lem01}
d_{T_\alpha^{(1)}}G^\alpha(j^2_x\phi)=\frac{\partial(G^\alpha\circ j^1\phi)}{\partial x^\alpha}\Big\vert_{x}
\end{equation}
for any $j^2_x\phi\in J^2\pi$.

  Since $X$ is locally given  by
 $$X=X^i(x,q,v) \frac{\partial }{\partial q^i}\circ \pi_{1,0}
$$
then $
X\circ \pi_{2,1}$ is locally given by
$$
X\circ \pi_{2,1}=\left(X^i(x,q,v) \frac{\partial }{\partial q^i}\circ \pi_{1,0}\right) \circ \pi_{2,1}
=X^i(x,q,v)  \circ \pi_{2,1} \,  \frac{\partial }{\partial q^i}\circ \pi_{2,0} \, .
$$

  From the above  local expressions of $X\circ \pi_{2,1}$ and the local expression (\ref{locdeltal}) of $ (\delta L)^V$ we obtain
\begin{equation}\label{lem02}
-(\delta L)^V(X\circ \pi_{2,1})(j^2_x\phi)= - \left( \frac{\partial }{\partial x^\alpha}\Big\vert_x\left(\frac{\partial L}{\partial v^i_\alpha}\circ j^1\phi \right)-\frac{\partial L}{\partial q^i}\Big\vert_{j^1_x\phi}\right) X^i (j^1_x\phi)
\end{equation}
  Now from (\ref{lem01}) and (\ref{lem02}) we obtain that if $\phi$ is
  a solution to the Euler-Lagrange equations then
   $$ \frac{\partial (G^\alpha\circ j^1\phi)}{\partial x^\alpha}\Big\vert_{x}=0\, .$$

   \item A direct computation using (\ref{locz1v}),
     (\ref{thetzeta40}),  (\ref{locdeltal}) and (\ref{locthetahat})
     proves the identity (\ref{iden00}).  \rule{5pt}{5pt}
   \end{enumerate}
  
Some classes of symmetries depend only on the variables  (coordinates)  in $E$. In this
section we  consider $v^i_\alpha$-dependent infinitesimal transformations which  can be regarded as vector fields $X$
along $\pi_{1,0}$. The following definition   is motivated by  Proposition \ref{600}.

\begin{definition}{Definition}
  A $\pi$-vertical vector field $X$ along $\pi_{1,0}$ is called a (generalized) symmetry if
  there exists a map $ (F^1, \ldots , F^k)\colon J^1\pi \to \mathbb{R}^k $ such that
\begin{equation}\label{luc00}
\,\,  d_{X^{(1)}}L(j^2_x\phi)=d_{T^{(1)}_\alpha}F^\alpha (j^2_x\phi)\,\,
\end{equation}
 for every solution $\phi$ to the Euler-Lagrange equations.
\end{definition}

The following version  of Noether's Theorem associates to each symmetry of the Lagrangian, in the sense given above, {\it  a conservation law}. \
\begin{theorem}{Theorem}\label{coche}
Let $X$ be a   symmetry of the Lagrangian $L$ then the  map $$ {\it G}=(G^1, \ldots , G^k)\colon J^1\pi \to \mathbb{R}^k $$ given by
 $$\,\, G^\alpha=F^\alpha- (\Theta_L^\alpha)^V(X)\,\,$$
defines a conservation law.
\end{theorem}

Proof:
  Let $X$ be a   symmetry of the Lagrangian $L$. Then from (\ref{iden00})
we get
$$
d_{T_\alpha^{(1)}}\left[ F^\alpha  - (\Theta_L^\alpha)^V(X)   \right]= -(\delta L)^V(X\circ \pi_{2,1})
$$
 and from Lemma \ref{lemsimgen}, the functions $G^\alpha=F^\alpha- (\Theta_L^\alpha)^V(X)$ define a conservation law.  \rule{5pt}{5pt}

\begin{example}{Example} Consider the homogeneous isotropic $2$-dimensional wave equation
\begin{equation}\label{waveeq}
\partial_{11}\phi -c^2\partial_{22}\phi-c^2\partial_{33}\phi=0,\end{equation}
where $\phi:\mathbb{R}^3 \to \mathbb{R}$ is a solution, and defines a section of the   trivial bundle $\pi \colon E=\mathbb{R}^3\times \mathbb{R} \to \mathbb{R}^3$.

Since  $J^1\pi=\mathbb{R}^3\times  T^1_3\mathbb{R}$,
equation (\ref{waveeq}) can be described as the Euler-Lagrange equation for the Lagrangian $ L \colon \mathbb{R}^3\times  T^1_3\mathbb{R} \to \mathbb{R}$ given by
$$ L(x,q,v)=\frac{1}{2}\left((v_1)^2-c^2(v_2)^2- c^2(v_3)^2\right).$$
In this case, for simplicity we consider the case $c=1$.

With the vector field $X=v_1 \displaystyle\frac{\partial }{\partial q}\circ\pi_{1,0}$   along $ \pi_{1,0}$ and the   functions   on $J^1\pi$
$$F^1(v_1,v_2,v_3) = -c^2(v_2)^2-c^2 (v_3)^2, \ \   F^2(v_1, v_2, v_3)= c^2 v_1v_2, \ \  F^3(v_1, v_2, v_3)=c^2v_1v_3$$
 using Theorem \ref{coche},  we deduce that   the following functions
  $$\begin{array}{ccccl}
  G^1&=&F^1- (\Theta^1_L)^V(X)&=&-c^2(v_2)^2-c^2 (v_3)^2-(v_1)^2 \\ \noalign{\medskip}
  G^2&=&F^2-(\Theta^2_L)^V(X)&=&2 c^2 v_1v_2 \\ \noalign{\medskip}
  G^3&=&F^3-(\Theta^3_L)^V(X)&=& 2c^2v_1v_3
  \end{array}$$
  define a conservation law.
\end{example}

\subsection{Variational symmetries}\label{sim-var}

In this section we consider the trivial bundle $\pi:E=\mathbb{R}^k \times Q \to \mathbb{R}^k $, and we recall some results of variational
 symmetries of the Euler-Lagrange equations that can be found in  Olver's book \cite{Olver}.

Let us remember that the solution of the Euler-Lagrange equation (\ref{e-l-2}) can be obtained as the extremals of the functional
\[
    \mathcal{L}(\phi)=\int_{\Omega_0}(L\circ j^1\phi)(x)d^kx\, ,
\]
where $d^kx=dx^1\wedge\cdots\wedge dx^k$ is the volume form on $\mathbb{R}^k $.
Roughly speaking, a variational symmetry is  a diffeomorphism that leaves the variational integral $\mathcal{L}$ unchanged.
\begin{definition}{Definition}\
\begin{enumerate}
\item  A {\it variational symmetry } is a diffeomorphism $\Phi\colon E=\mathbb{R}^k \times Q\to E=\mathbb{R}^k \times Q$ verifying the following conditions:
 \begin{enumerate}
    \item It is a fiber-preserving map for the bundle $\pi\colon E\to \mathbb{R}^k $; that is, $\Phi$ induces a diffeomorphisms $\varphi\colon \mathbb{R}^k \to \mathbb{R}^k $ such that $\pi\circ \Phi=\varphi\circ \pi$
    \item If $\tilde{x}=\varphi(x)$ for each $x\in \mathbb{R}^k $
        \[
            \int_{\tilde{\Omega} }(L\circ j^1(\Phi\circ \phi\circ \varphi^{-1} ))(\tilde{x})d^k\tilde{x}= \int_{\Omega}(L\circ j^1\phi)(x)d^kx\,
        \] where $\tilde{\Omega}=\varphi(\Omega)$.
 \end{enumerate}
\item An infinitesimal variational symmetry is a vector field $X\in \mathfrak{X}(\mathbb{R}^k \times Q)$ whose local flows are variational symmetries.
\end{enumerate}
\end{definition}

The following results can be seen in Theorem $4.12$ and in Corollary $4.30$ \cite{Olver} .

\begin{theorem}{Theorem}\label{consolver0}
i) A vector field $X$ on $\mathbb{R}^k \times Q$  is  a variational symmetry if, and only if,
$X^1(L)+ L \, d_{T_\alpha^{(0)}}X_\alpha  =0$, where $X_\alpha=dx^\alpha(X)$.

\noindent ii) If    $X$ is a variational symmetry then
 $ \Theta^\alpha_L(X^1)$  defines a conservation law.
\end{theorem}

 \begin{example}{Example} We consider again  the homogeneous isotropic $2$-dimensional wave equation (\ref{waveeq}).

 The rotation group $X=-x^3\displaystyle\frac{\partial }{\partial x^2}+x^2\displaystyle\frac{\partial }{\partial x^3}$ is a variational symmetry, and then the corresponding conservation law $(\Theta^1_L(X^1),\Theta^2_L(X^1), \Theta^3_L(X^1))$ is given by the functions
$$
\left(x^3v_1v_2-x^2v_1v_3, \, -\frac{1}{2} x^3 \, u +x^2v_2v_3, \,  -\frac{1}{2} x^2\, u-v_3v_2x^3\right)
$$
where $u= (v_1)^2+(v_2)^2+(v_3)^2$.
\end{example}

 \begin{example}{Example}  We consider again $Q=\mathbb{R}$, and let
$$ 0=\left(1+(\partial_2\phi)^2\right)\partial_{11}\phi-2\partial_1\phi\, \partial_2\phi\, \partial_{12}\phi+\left(1+(\partial_1\phi)^2\right)\partial_{22}\phi $$
be  the equation of minimal surfaces, which is the Euler-Lagrange equations for the Lagrangian $L \colon \mathbb{R}^2\times T^1_2\mathbb{R} \to \mathbb{R}$ defined by $L(x^1,x^2,q,v_1,v_2)=\sqrt{1+(v_1)^2+(v_2)^2}$. The vector field $X=-q\displaystyle\frac{\partial }{\partial x^1}-q\displaystyle\frac{\partial }{\partial x^2}+(x^1+x^2)\displaystyle\frac{\partial }{\partial q}$ is a variational symmetry, and then the corresponding conservation law
$(\Theta^1_L(X^1),\Theta^2_L(X^1))$ is given by the functions
 $$
 \left(\frac{-q(1+(v_2)^2-v_1v_2)+(x^1+x^2)v_1}{\sqrt{1+(v_1)^2+(v_2)^2}},\, \, \frac{-q(1+(v_1)^2-v_1v_2)+(x^1+x^2)v_2}{\sqrt{1+(v_1)^2+(v_2)^2}}\right)
 $$
\end{example}

Now we  describe some relationships between the above symmetries.
\begin{theorem}{Theorem}
i) Let $X$ be a $\pi$-vertical variational symmetry. Then the vector field $ X\circ \pi_{1,0}$ along $\pi_{1,0}$   is a generalized symmetry.

ii) The conservation law induced by $X$ and $ X\circ \pi_{1,0}$  coincide.
\end{theorem}

Proof:
i) Since $X $ is a $\pi$-vertical variational symmetry, then locally $X=X^i(x,q)\frac{\partial }{\partial q^i}$, and from Theorem \ref{consolver0} we know that $X^1(L)=0$. From (\ref{xx}) we know that  $(X\circ \pi_{1,0})^{(1)}=X^1\circ \pi_{2,1}$. Then,
$$ d_{(X\circ \pi_{1,0})^{(1)}}L(j^2_x\phi)=(X\circ \pi_{1,0})^{(1)}(j^2_x \phi)(L)=X^1(j^1_x\phi)(L)=0$$
and thus $X$ is a generalized symmetry.

ii) It is a consequence of
$\Theta^\alpha_L(X^1)= (\Theta_L^\alpha)^V(X\circ \pi_{1,0}).$
 \rule{5pt}{5pt}

\subsection{Noether symmetries}\label{sim-noe}

%

  In the paper \cite{mrsv}  we  introduced the following
  definition:
  \begin{definition}{Definition}
 A vector field $Y \in \mathfrak{X}(\mathbb{R}^k\times T^1_kQ)$ is an infinitesimal Noether symmetry
if
$$
\mathcal{L}_Y\omega_L^\alpha=0, \quad i_Y dx^\alpha=0, \quad
\mathcal{L}_Y E_L=0.
$$
\end{definition}

\begin{theorem}{Theorem}
Let $X$ be a $\pi$-vertical vector field  on $\mathbb{R}^k\times Q$ such that $X^1$ is
an infinitesimal Noether symmetry, then $X=X\circ\pi_{1,0}$ is a generalized symmetry.
\end{theorem}

Proof:
Using (\ref{locX1gen}), (\ref{LagranEner}), then from the local
expression $X =X^i\displaystyle\frac{\partial }{\partial q^i} $ and the condition $\mathcal{L}_{X^1}E_L=0$, we deduce that \begin{equation}\label{yy}(X^i\circ\pi_{1,0})\frac{\partial L}{\partial q^i}= v^i_\alpha X^1\left(\frac{\partial L}{\partial v^i_\alpha}\right)\quad .\end{equation}
From the condition   $\mathcal{L}_{X^1}\omega^\alpha_L=0$ we obtain $d(\mathcal{L}_{X^1}\theta^\alpha_L)=0$ and so    there exist (locally defined)  functions $F^\alpha \colon U\subset\mathbb{R}^k\times T^1_kQ \to \mathbb{R}$ such that $$\mathcal{L}_{X^1}\theta^\alpha_L=dF^\alpha \,
\quad 1\leq \alpha \leq k \, . $$
With these identities we obtain the following relations
\begin{equation}\label{fat}
\begin{array}{ll}
\displaystyle\frac{\partial F^\alpha}{\partial x^\beta}=\displaystyle\frac{\partial L}{\partial v^i_\alpha}\displaystyle\frac{\partial X^i}{\partial x^\beta}\circ \pi_{1,0} &
\displaystyle\frac{\partial F^\alpha}{\partial q^j}=\displaystyle\frac{\partial L}{\partial v^i_\alpha}\displaystyle\frac{\partial X^i}{\partial q^j}\circ \pi_{1,0}-X^1\left(\displaystyle\frac{\partial L}{\partial v^j_\alpha}\right) \\ \noalign{\bigskip}
\displaystyle\frac{\partial F^\alpha}{\partial v^j_\beta}=\displaystyle\frac{\partial L}{\partial v^i_\alpha}\displaystyle\frac{\partial X^i}{\partial v^j_\beta}\circ \pi_{1,0}=0 &
 \end{array}\end{equation}

 From (\ref{xx}), (\ref{yy})  and (\ref{fat}),  we deduce that
$$\begin{array}{ccl}
d_{(X\circ\pi_{1,0})^{(1)}}L(j^2_x\phi)&=&X^i(\phi(x))\displaystyle\frac{\partial L}{\partial q^i}\Big\vert_{j^1_x\phi}+\left(\displaystyle \frac{\partial X^i}{\partial x^\alpha}\Big\vert_{\phi(x)}+v^j_\alpha(j^1_x\phi)\displaystyle\frac{\partial X^i}{\partial q^j}\Big\vert_{j^1_x\phi}\right)\displaystyle\frac{\partial L}{\partial v^i_\alpha}\Big\vert_{j^1_x\phi} \\ \noalign{\medskip}
&=& X^i(\phi(x))\displaystyle\frac{\partial L}{\partial q^i}\Big\vert_{j^1_x\phi}+\displaystyle\frac{\partial F^\alpha}{\partial x^\alpha}\Big\vert_{j^1_x\phi}  + v^j_\alpha(j^1_x\phi)\left(\displaystyle\frac{\partial F^\alpha}{\partial q^j}\Big\vert_{j^1_x\phi} +X^1(j^1_x\phi)\left(\displaystyle\frac{\partial L}{\partial v^j_\alpha}\right)\right) \\ \noalign{\medskip}
&=& \displaystyle\frac{\partial F^\alpha}{\partial x^\alpha}\Big\vert_{j^1_x\phi} +v^j\alpha({j^1_x\phi})\displaystyle\frac{\partial F^\alpha}{\partial q^j}\Big\vert_{j^1_x\phi}=d_{T^{(1)}_\alpha}F^\alpha(j^2_x\phi)
\end{array}$$
for any $j^2_x\phi$.

This proves that $X\circ\pi_{1,0}$ is a generalized symmetry. \rule{5pt}{5pt}

\section*{Conclusions}
In this paper, we have discussed  a new geometric formalism for fiber bundles over Euclidean spaces; this new formalism
allows us to understand better the similarities and differences between the multisymplectic and $k$-cosymplectic settings. Even if such a fiber bundle is topologically trivial, it has some interest from the geometric point of view. Indeed, it is a way to understand better the multisymplectic formalism (and, by the way, it is a usual case in Continuun Mechanics). The current paper is a first step to get more treatable ways to work with the field equations when the base space (the space-time manifold in the physical contexts) is not trivial or even a parallelizable manifold. We are considering more general situations and this paper will help very much for more generalizations.

\section*{Acknowledgments}
We acknowledge the financial support of the Ministerio de Ciencia e Innovaci\'{o}n (Spain), projects MTM2011-22585, MTM2011-15725-E, MTM2010-21186-C02-01,  the European project IRSES-project Geomech-246981 and the ICMAT Severo Ochoa project SEV-2011-0087.

\end{document}